\documentclass[10pt,journal,compsoc]{IEEEtran}
\usepackage{epsfig}
\usepackage{graphicx}
\usepackage{caption}
\usepackage{subcaption}
\usepackage{amsmath}
\usepackage{amssymb,amsfonts}
\usepackage{mathrsfs} 
\usepackage{enumerate}
\usepackage{multirow}
\usepackage{tabularx}
\usepackage{float}
\usepackage{color}
\usepackage{times}
\usepackage{comment}
\usepackage{makecell}
\usepackage{bbm}
\usepackage{url}

\DeclareMathOperator*{\argmin}{argmin}

\makeatletter
\def\hlinewd#1{%
\noalign{\ifnum0=`}\fi\hrule \@height #1 %
\futurelet\reserved@a\@xhline} 
\makeatother

\begin{document}
\title{End-to-End Signal Classification in Signed Cumulative Distribution Transform Space}

\author{Abu Hasnat Mohammad Rubaiyat, 
    Shiying Li, 
    Xuwang Yin, 
    Mohammad Shifat-E-Rabbi, 
    Yan Zhuang, 
    and~Gustavo K. Rohde%
\thanks{A. H. M. Rubaiyat, X. Yin,  and Y. Zhuang are with the Department of Electrical and Computer Engineering, University of Virginia, Charlottesville, VA 22904 USA (e-mail: \{ar3fx, xy4cm, yz8bk\}@virginia.edu).}%
\thanks{S. Li and M. Shifat-E-Rabbi are with the Department of Biomedical Engineering, University of Virginia, Charlottesville, VA 22908, USA (e-mail: \{sl8jx, mr2kz\}@virginia.edu).}%
\thanks{G. K. Rohde is with the Department of Biomedical Engineering and the Department of Electrical and Computer Engineering, University of Virginia, Charlottesville, VA 22908, USA (e-mail: gustavo@virginia.edu).}%
}

\IEEEtitleabstractindextext{%
\begin{abstract}
This paper presents a new end-to-end signal classification method using the signed cumulative distribution transform (SCDT). We adopt a transport-based generative model to define the classification problem. We then make use of mathematical properties of the SCDT to render the problem easier in transform domain, and solve for the class of an unknown sample using a nearest local subspace (NLS) search algorithm in SCDT domain. Experiments show that the proposed method provides high accuracy classification results while being computationally cheap, data efficient, and robust to out-of-distribution samples with respect to the existing end-to-end classification methods. The implementation of the proposed method in Python language is integrated as a part of the software package PyTransKit \cite{pytranskit}.
\end{abstract}

}
 
\maketitle

\IEEEdisplaynontitleabstractindextext

\IEEEpeerreviewmaketitle

\IEEEraisesectionheading{\section{Introduction}\label{sec:intro}}
\IEEEPARstart{S}{ignal} (time series) classification is considered a challenging problem in data science. It refers to the automatic prediction of the class label of an unknown 
time series event using the information extracted from the corresponding signal intensities. Time series data classification tasks can be found in many applications such as human activity recognition (HAR) \cite{lara2012survey}, physiological signal assessment \cite{berkaya2018survey}\cite{subasi2010eeg}, communications \cite{fehske2005new}, structural or machine health monitoring systems \cite{abdeljaber20181} \cite{zhao2019deep}, financial modeling \cite{lu2009financial}, and others. In many applications (e.g., HAR, ECG, etc.) the time series events of interest can be modeled as instances of a certain (often unknown) template or prototype pattern observed under unknown time warps \cite{rubaiyat2022nearest}. Here we propose a new end-to-end tool for classification of signals or time series events of this type using the signed cumulative distribution transform (SCDT), a new mathematical signal transform introduced in \cite{aldroubi2022signed}.

Existing signal classification approaches can be categorized into two broad, and at times overlapping, categories: 1) feature engineering-based classifiers and 2) end-to-end learning classifiers, such as convolutional neural networks (CNNs). Feature engineering-based methods \cite{nanopoulos2001feature} \cite{bagnall2017great} \cite{fulcher2014highly} usually rely on the extraction of numerical features (e.g., time domain features, frequency domain features, wavelet features) from the raw signal data, and then the application of different multivariate regression-based classification methods including linear discriminant analysis, support vector machines, random forests, and others. Deep learning-based signal classification methods \cite{fawaz2019deep} \cite{wang2019deep} \cite{al2016deep}, on the other hand, connect the raw input data to the output class label by utilizing a large number of hidden layers. These methods have widely been studied recently as they have shown high accuracy in certain classification tasks. 

Feature engineering signal classification approaches primarily differ in the types of features chosen to characterize each signal. The bag-of-features framework \cite{baydogan2013bag} extracts interval features using fixed- and variable-length intervals, and trains a classifier on the extracted features. 
Ensemble-based approaches such as COTE \cite{bagnall2015time}, HIVE-COTE \cite{lines2018time}, time series forest (TSF) \cite{deng2013time}, and others combine different features and classifiers to achieve high classification accuracy. Most of these methods require crafting especially designed features (feature engineering) as well as some amount of data preprocessing.


Traditional end-to-end signal classification techniques include distance-based methods \cite{abanda2019review} \cite{lines2015time} that work directly on raw time series with some similarity measures such as Euclidean distance or dynamic time warping (DTW) \cite{berndt1994using}. Particularly, a combination of $1$-nearest neighbor (1NN) with DTW distance is known to be a very effective time series classification approach \cite{xi2006fast}. However, it is known to have high computational complexity \cite{zheng2014time}.
Approaches based on deep neural networks, especially convolutional neural networks (CNN) \cite{zhao2017convolutional}, have been explored in recent years for end-to-end signal classification. Wang et al. \cite{wang2017time}, for example, provided three standard deep learning benchmark models for time series classification: deep multilayer perceptrons (MLP), fully convolutional networks (FCN), and residual networks (ResNet). The method known as multi-scale convolutional neural network (MCNN) \cite{cui2016multi} is another deep learning approach that takes advantage of CNNs for end-to-end classification of univariate time series. Karim et al. \cite{karim2017lstm} proposed the LSTM-FCN, an improvement over FCN by augmenting the FCN module with a Long Short Term Recurrent Neural Network (LSTM RNN) sub-module. Though they can produce accurate results in many instances, these methods tend to require extensive amounts of training data, are computationally expensive, and often vulnerable to out-of-distribution examples. Furthermore, existing end-to-end classifiers often lack a proper data model and a clear formulation of the classification problem, making the classification models difficult to interpret. In particular, the lack of an underlying mathematical foundation often leads to uncertainty as to which exact situations or applications they will work, and when they will fail.


In recent years, some effort has been made to exploit transport transforms for signal classification \cite{kolouri2017optimal} as an alternative to the techniques mentioned above. The cumulative distribution transform (CDT), based on the 1D Wasserstein embedding, was introduced in \cite{Park:18} as a means of classifying strictly positive signals following the linear optimal transport framework proposed in \cite{wang2013linear}. Aldroubi et al. \cite{aldroubi2022signed} extended the CDT to general signed signals and proposed the signed cumulative distribution transform (SCDT). 
Both transforms have a number of properties which allow one to solve nonlinear classification problems using linear classifiers, such as Fisher discriminant analysis, support vector machines, or logistic regression, in signal transform space. The classification method proposed in \cite{rubaiyat2022nearest} utilizes the nearest subspace method to classify signals in SCDT domain. Assuming the data corresponding to a particular class as compositions of a single template, this method formed a linear subspace for each class. The method we propose here is an extension of this approach.

In this paper, we propose a new generative model to represent the signal data such that the signals from each class can be seen as observations of a set of unknown templates under some unknown deformations. We then formulate a signal classification problem for the data that follows the generative model, and employ a nearest local subspace search algorithm in SCDT domain to devise a solution. We demonstrate the advantages of the method over state-of-the-art deep learning methods in terms of classification accuracy on a 10 datasets with comprehensive experiments. The proposed method also provides competitive performance in classifying signals with very low computational cost with respect to a distance based end-to-end system (1NN-DTW). In addition, experiments highlight other interesting properties of the method in comparison to alternative end-to-end solutions including data efficiency and robustness to out-of-distribution conditions.


The remaining of this paper is organized as follows: in section \ref{sec:prel}, we briefly review the definitions and properties of the CDT and the SCDT. In section \ref{sec:method}, we state the classification problem and the proposed solution. Experimental setup, datasets, and results are described in section \ref{sec:exp}, with the discussion of the results in section \ref{sec:diss}. Finally, section \ref{sec:conc} provides  concluding remarks.

\section{Preliminaries}\label{sec:prel}
\subsection{Notation}
Throughout the manuscript, we work with $L_1$ signals $s$, i.e. $\int_{\Omega_s}|s(t)|dt<\infty$, where $\Omega_s\subseteq\mathbb{R}$ is the domain over which $s$ is defined. We use $s^{(c)}_{j,m}$ to represent a signal generated from the $m$-th template of class $c$ under deformation $g_j$.
We denote the $m$-th template from class $c$ as $\varphi_m^{(c)}$. Some symbols used throughout the manuscript are listed in Table \ref{tab:symbols}.
\begin{table}[bt]
    \centering
    \normalsize
    \caption{Description of symbols}
        \begin{tabular}{ll}
        \hline
        Symbols                & Description    \\ \hline
        $s(t)$ & Signal \\
        $s_0(y)$ & Reference signal to calculate the transform\\
        $\widehat{s}(y)$ & \makecell[l]{SCDT of signal $s(t)$}\\
        $g(t)$ & Strictly increasing and differentiable function \\
        $s\circ g$ & $s(g(t))$: composition of $s(t)$ with $g(t)$\\
        $\mathcal{T}$ & Set of all possible increasing diffeomorphisms\\
        $\mathbb{S}/\widehat{\mathbb{S}}$ & Set of signals$/$SCDT of the signals \\
        \hline
        \end{tabular}
    \label{tab:symbols}
\end{table}

\subsection{The Cumulative Distribution Transform}
The Cumulative Distribution Transform (CDT) was introduced in \cite{Park:18} for positive smooth normalized functions. It is an invertible nonlinear 1D signal transform from the space of smooth positive probability densities to the space of diffeomorphisms, which can be described as follows: let $s(t), t\in\Omega_s$ and $s_0(y),y\in\Omega_{s_0}$ define a given signal and a reference signal, respectively, such that $\int_{\Omega_s}s(u)du = \int_{\Omega_{s_0}}s_0(u)du = 1$ and $s_0(y), s(t)>0$ in their respective domains. The CDT of the signal $s(t)$ is then defined to be the function $s^*(y)$ that solves,
\begin{align}
    \int_{\inf(\Omega_s)}^{s^*(y)} s(u)du = \int_{\inf(\Omega_{s_0})}^{y} s_0(u)du.
    \label{eq:cdt}
\end{align}
Now considering the cumulative distribution functions (CDFs) $S(t) = \int_{-\infty}^{t} s(u)du$ and $S_0(y) = \int_{-\infty}^{y} s_0(u)du$, an alternative expression for $s^*(y)$ is given by,
\begin{equation}
	s^*(y) = S^{-1}(S_0(y)).
	\label{eq:cdt_alt}
\end{equation}
The CDT is therefore seen to inherit the domain of the reference signal. Moreover, if the uniform reference signal is used (i.e., $s_0(y) = 1$ in $\Omega_{s_0}=[0,1]$), we can write $S_0(y) = y$ and $s^*(y) = S^{-1}(y)$. That is to say, the CDT is the inverse of the cumulative distribution function of the given signal $s(t)$. Note that the definition of the CDT described above is slightly different from the formulation used in \cite{Park:18}. For simplicity, here we use the CDT definition described in \cite{Rubaiyat:20}. The CDT is invertible, and the inverse formula is defined in differential form as:
\begin{equation}
	s(t) = \left(s^{*^{-1}}(t)\right)' s_0( s^{*^{-1}}(t)).
\end{equation}
Although the CDT can be used in solving many classification \cite{Park:18} and estimation \cite{Rubaiyat:20} problems, the framework described above is defined only for positive density functions. Aldroubi et al. \cite{aldroubi2022signed} extended the CDT to general finite signed signals and named the new signal transformation technique as the signed cumulative distribution transform (SCDT).

\subsection{The Signed Cumulative Distribution Transform}
The signed cumulative distribution transform (SCDT) \cite{aldroubi2022signed} is an extension of the CDT, which is defined for general finite signed signals with no requirements on the total mass. First, the transform is defined for the non-negative signal $s(t)$ with arbitrary mass as:
\begin{equation}
    \widehat{s}(y) = \begin{cases}
    \left(s^*(y),\|s\|_{L_1}\right),& \text{if } s\neq 0\\
    (0,0),              & \text{if } s=0,
\end{cases}
\label{eq:scdt_mass}
\end{equation}
where $\|s\|_{L_1}$ is the $L_1$ norm of signal $s$ and $s^*$ is the CDT (defined in eqn. (\ref{eq:cdt_alt})) of the normalized signal $\frac{1}{\|s\|_{L_1}}s$.

\begin{figure}[tb]
    \centering
    \includegraphics[width=0.43\textwidth]{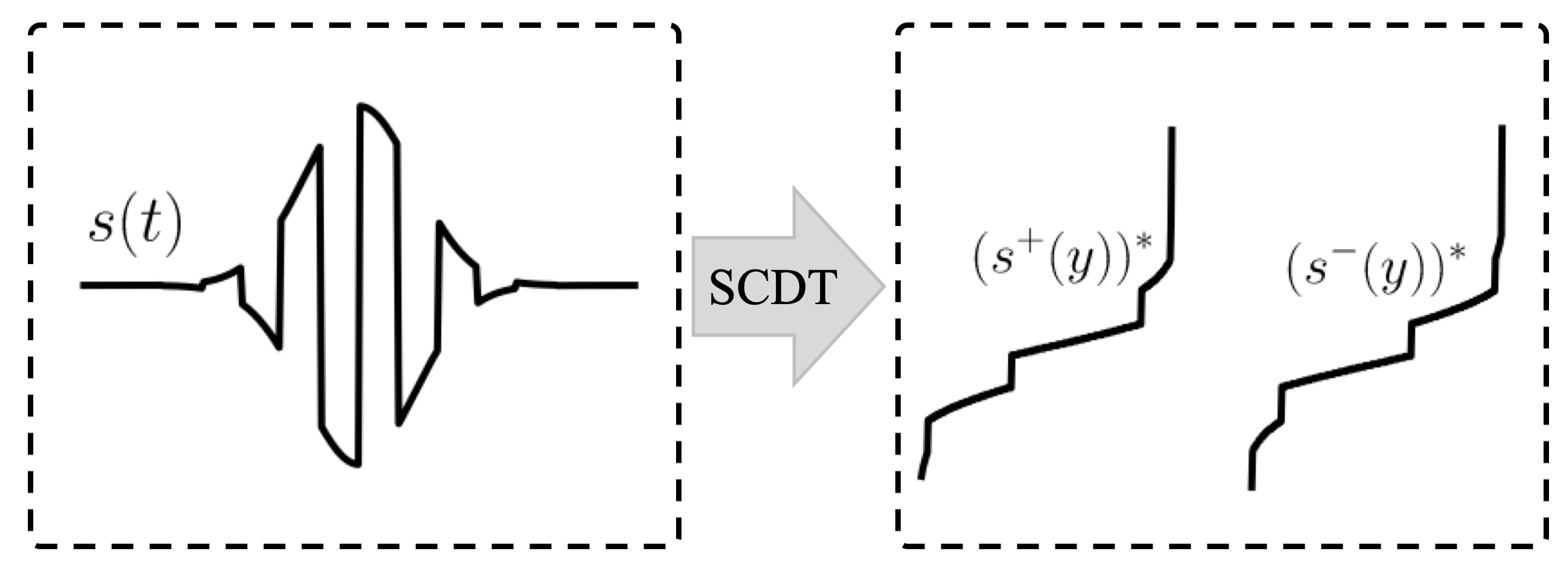}
    \caption{SCDT (without the constant terms) of an example signal. 
    }
    \label{fig:scdt}
    \vspace{-1em}
\end{figure}

Now for a signed signal, the Jordan decomposition \cite{royden1988real} is used to define the transform. The Jordan decomposition of a signed signal $s(t)$ is given by $s(t) = s^+(t) - s^-(t)$, where $s^+(t)$ and $s^-(t)$ are the absolute values of the positive and negative parts of the signal $s(t)$. The SCDT of $s(t)$ is then defined as:
\begin{equation}
    \widehat{s}(y) = \left(\widehat{s}^+(y), \widehat{s}^-(y)\right),
    \label{eq:scdt}
\end{equation}
where $\widehat{s}^+(y)$ and $\widehat{s}^-(y)$ are the transforms (defined in eqn. (\ref{eq:scdt_mass})) for the signals $s^+(t)$ and $s^-(t)$, respectively. Fig. \ref{fig:scdt} shows an example of the SCDT of a signal. Like the CDT, the SCDT is also an invertible operation, with the inverse being,
\begin{align}
    s(t) = &\|s^{+}\|_{L_1}\left((s^{+})^{*^{-1}}(t)\right)'s_0( (s^{+})^{*^{-1}}(t)) \nonumber\\ 
    &- \|s^{-}\|_{L_1}\left((s^{-})^{*^{-1}}(t)\right)'s_0( (s^{-})^{*^{-1}}(t)).
\end{align}
Moreover, the SCDT has a number of properties that will help us simplify the signal classification problems.

\subsubsection{Composition property}
If the SCDT of a signed signal $s$ is denoted as $\widehat{s}$, the SCDT of the signal $s_g=g's\circ g$ is given by:
\begin{equation}
    \widehat{s}_g = \left(g^{-1}\circ (s^+)^*,\|s^{+}\|_{L_1},g^{-1}\circ (s^-)^*,\|s^{-}\|_{L_1}\right),
    \label{eq:scdt_composition}
\end{equation}
where, $g(t)$ is an invertible and differentiable increasing function, $s\circ g = s(g(t))$, and $g'(t)=dg(t)/dt$ \cite{aldroubi2022signed}. For example, in case of shift and linear dispersion (i.e., $g(t) = \omega t - \tau$) of a given signal $s(t)$, the SCDT of the signal $s_g(t)=\omega s(\omega t - \mu)$ can be derived from composition property as:
\begin{equation*}
    \widehat{s}_g = \left(\frac{(s^+)^* + \mu}{\omega},\|s^{+}\|_{L_1},\frac{(s^-)^* + \mu}{\omega},\|s^{-}\|_{L_1}\right).
\end{equation*}
The composition property implies that variations along the independent variable caused by $g(t)$ will change only the dependent variable in the transform domain. 

\subsubsection{Convexity property}
\label{sec:convexity}
Let $\mathbb{S}=\{s_j|s_j=g'_j \varphi\circ g_j, \forall g_j\in \mathcal{G}\}$ be a set of signals, where $\varphi$ is a given signal and $\mathcal{G}\subset \mathcal{T}$ denotes a set of 1D spatial deformations of a specific kind (e.g., translation, dilation, etc.). The convexity property of the SCDT \cite{aldroubi2022signed} states that the set $\widehat{\mathbb{S}} = \{\widehat{s}_j:s_j\in \mathbb{S}\}$ is convex for every $\varphi$ if and only if $\mathcal{G}^{-1}=\{g_j^{-1}:g_j\in \mathcal{G}\}$ is convex.

The set $\mathbb{S}$ defined above can be interpreted as a generative model for a signal class while $\varphi$ being the template signal corresponding to that class. In the next section we describe a generative model-based problem formulation for time series event classification, and then show how the composition and convexity properties of the SCDT help facilitate signal classification.

\section{Proposed Method}\label{sec:method}
\subsection{Generative model and problem statement}\label{sec:problem}
In \cite{rubaiyat2022nearest}, a generative model-based problem statement was proposed for signal classification problems where each signal class can be modeled as instances of a certain template observed under some unknown deformations (e.g., translation, dispersion, etc.). Signal classes of such type can be described with the following generative model:

\subsubsection{Generative model (single template)} 
Let $\mathcal{G}^{(c)}\subset \mathcal{T}$ denote a set of increasing 1D deformations of a specific kind, where $\mathcal{T}$ is a set of all possible increasing diffeomorphisms from $\mathbb{R}$ to $\mathbb{R}$. The 1D mass preserving generative model for class $c$ is then defined to be the set:
\begin{align}
    \mathbb{S}^{(c)}=\{s_j^{(c)}|s_j^{(c)}=g'_j\varphi^{(c)}\circ g_j, g_j\in \mathcal{G}^{(c)}, g'_j>0\},
    \label{eq:genmod_single}
\end{align}
where $s_j^{(c)}$ is the $j$-th signal from class $c$, and $\varphi^{(c)}$ is the template pattern corresponding to that class. However, in many applications it is difficult to find a signal class that can be represented using the single template-based generative model defined above. In this paper, we use a multiple template-based generative model to represent such signal classes.

\begin{figure}[tb]
    \centering
    \includegraphics[width=0.44\textwidth]{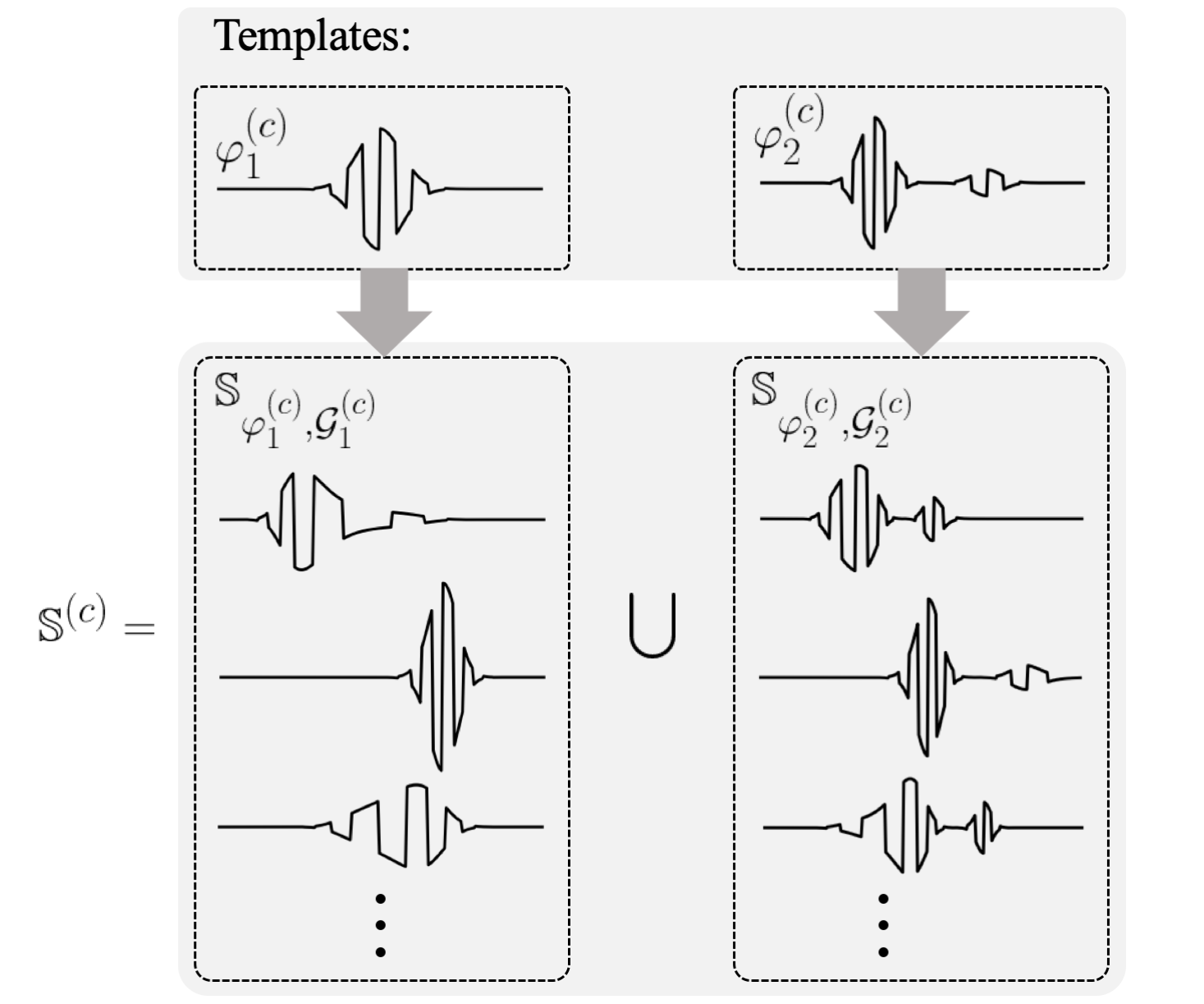}
    \caption{Generative model example for a signal class `$c$'. The set of all signals from class $c$ is denoted as $\mathbb{S}^{(c)}$, which is modeled as the union of subsets $\mathbb{S}_{\varphi_m^{(c)},\mathcal{G}_m^{(c)}}$ (for $m=1,2$) containing data generated from a template $\varphi_m^{(c)}$ under deformation $\mathcal{G}_m^{(c)}$.}
    \label{fig:sig_deform}
\end{figure}

\subsubsection{Generative model (multiple templates)}
\begin{figure*}[htb]
    \centering
    \includegraphics[width=0.9\textwidth]{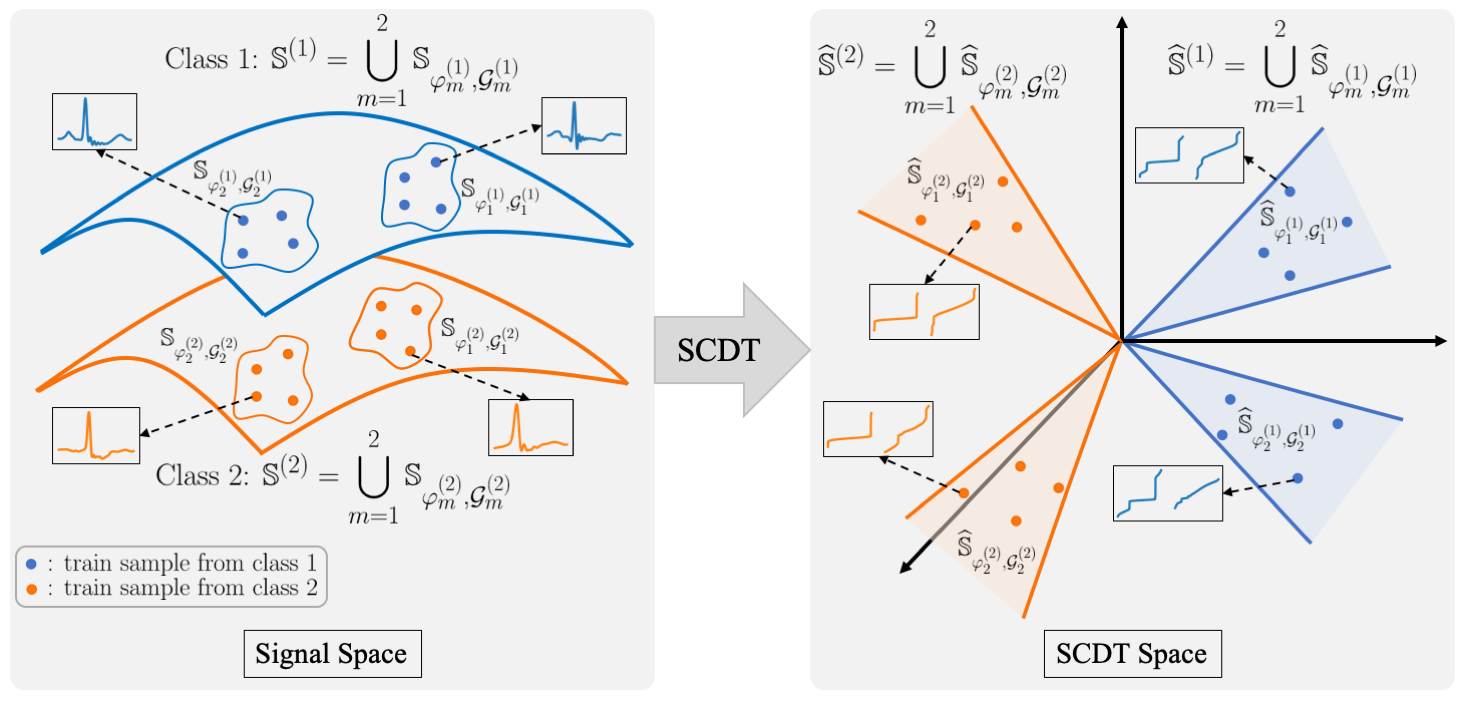}
    \caption{Geometric interpretation of data following the proposed generative model defined in equations (\ref{eq:genmod_multi}) and (\ref{eq:genmod_scdt}). On the left panel, two classes ($\mathbb{S}^{(1)} \text{and } \mathbb{S}^{(2)}$) are depicted in signal space. The set $\mathbb{S}^{(c)}$ for class $c$ is modeled as the union of two subsets: $\mathbb{S}_{\varphi_1^{(c)},\mathcal{G}_1^{(c)}}$ and $\mathbb{S}_{\varphi_2^{(c)},\mathcal{G}_2^{(c)}}$, $c=1,2$, where both the subsets are non-convex. The right panel shows the geometry of the signal classes in SCDT domain modeled as the union of convex (for convex $(\mathcal{G}_1^{(c)})^{-1}$ and $(\mathcal{G}_2^{(c)})^{-1}$ as defined in (\ref{eq:genmod_multi})) subsets: $\widehat{\mathbb{S}}_{\varphi_1^{(c)},\mathcal{G}_1^{(c)}}$ and $\widehat{\mathbb{S}}_{\varphi_2^{(c)},\mathcal{G}_2^{(c)}}$ for $c=1,2$.
    }
    \label{fig:data_geo}
\end{figure*}
For a set of increasing 1D spatial deformations denoted as $\mathcal{G}^{(c)}_m\subset\mathcal{T}$, the 1D mass preserving generative model for class $c$ is defined to be the set:
\begin{align}
    &\mathbb{S}^{(c)} = \bigcup\limits_{m=1}^{M_c} \mathbb{S}_{\varphi_m^{(c)},\mathcal{G}_m^{(c)}}, \nonumber\\
    &\mathbb{S}_{\varphi_m^{(c)},\mathcal{G}_m^{(c)}}=\left\{s_{j,m}^{(c)}|s_{j,m}^{(c)}=g'_{j}\varphi_m^{(c)}\circ g_{j}, g'_{j}>0, g_{j}\in \mathcal{G}_m^{(c)}\right\},\nonumber\\
    &\left(\mathcal{G}_m^{(c)}\right)^{-1}=\left\{\sum_{i=1}^k\alpha_i f_{i,m}^{(c)}, \alpha_i\geq 0\right\},
    \label{eq:genmod_multi}
\end{align}
where $\left\{f_{1,m}^{(c)}, f_{2,m}^{(c)},..., f_{k,m}^{(c)}\right\}$ denotes a set of linearly independent and strictly increasing (within the domain of the signals) functions.
Meaning the generative model for class $c$ is modeled as the union of $M_c$ subsets, where each subset ($\mathbb{S}_{\varphi_m^{(c)},\mathcal{G}_m^{(c)}}$) corresponds to data generated from a particular template ($\varphi_m^{(c)}$) under some time deformations ($\mathcal{G}_m^{(c)}$). Here, $M_c$ is the total number of templates used to represent class $c$, $\varphi_m^{(c)}$ is the $m$-th template signal from class $c$, and $s_{j,m}^{(c)}$ is the $j$-th signal generated from $m$-th template under deformation defined by $g_j$. 
Fig. \ref{fig:sig_deform} illustrates a few examples of such deformations. Considering the generative model, the classification problem can be defined as follows:\\

\textbf{Classification problem:} \textit{Let $\mathcal{G}^{(c)}_m\subset \mathcal{T}$ be a set of spatial deformations and $\mathbb{S}^{(c)}$ be defined as in eq. (\ref{eq:genmod_multi}), for classes $c=1, \cdots, N_c$. Given a set of training samples $\{s_{1}^{(c)}, s_{2}^{(c)}, ...\}\subset \mathbb{S}^{(c)}$ for class $c$, determine the class label of an unknown signal $s$}.

\subsection{Proposed solution}\label{sec:CDT_est}
We propose a solution to the classification problem defined above using the SCDT in combination with the nearest local subspace method. For a given test sample, the algorithm first searches for $k$ closest training samples (from a particular class) to the test signal in SCDT domain based on a distance definition specified later. 
Next, a local subspace is spanned by these samples for each class. The unknown class label of the test sample is then estimated based on the shortest distance from the SCDT of the test signal to these local subspaces.

As stated in \cite{rubaiyat2022nearest} and \cite{shifat2020radon}, the generative model described in eq. (\ref{eq:genmod_multi}) generally yields nonconvex signal classes, causing the above classification problem to be difficult to solve. As specified above in section \ref{sec:convexity} (see \cite{aldroubi2022signed} for more details), under certain assumptions, the geometry of the signal class can be simplified. Hence, the proposed solution begins with applying the SCDT defined in eq. (\ref{eq:scdt}) on the input signals. The generative model in the transform domain is then given by,
\begin{align}
    &\widehat{\mathbb{S}}^{(c)} = \bigcup\limits_{m=1}^{M_c} \widehat{\mathbb{S}}_{\varphi_m^{(c)},\mathcal{G}_m^{(c)}}, \nonumber\\
    &\widehat{\mathbb{S}}_{\varphi_m^{(c)},\mathcal{G}_m^{(c)}} = \left\{\widehat{s}_{j,m}^{(c)}| \widehat{s}_{j,m}^{(c)}=g_{j}^{-1}\circ \widehat{\varphi}_m^{(c)}, g_j\in \mathcal{G}_m^{(c)}\right\}, \nonumber\\
    &\left(\mathcal{G}_m^{(c)}\right)^{-1}=\left\{\sum_{i=1}^k\alpha_i f_{i,m}^{(c)}, \alpha_i\geq 0\right\},
    \label{eq:genmod_scdt}
\end{align}
where $g_{j}^{-1}\circ \widehat{\varphi}_m^{(c)}$ is the SCDT of the signal $g'_j\varphi_m^{(c)}\circ g_j$. Here, the set $\left(\mathcal{G}_m^{(c)}\right)^{-1}$ is convex by definition. Therefore, using the convexity property highlighted earlier, it can be shown that $\widehat{\mathbb{S}}_{\varphi_m^{(c)},\mathcal{G}_m^{(c)}}$ given in eq.~\eqref{eq:genmod_scdt} forms a convex set. Moreover, since the SCDT is a one-to-one map, it follows that if $\mathbb{S}_{\varphi_m^{(c)},\mathcal{G}_m^{(c)}} \cap \mathbb{S}_{\varphi_w^{(p)},\mathcal{G}_w^{(p)}}=\varnothing$ for $c\neq p$, then $\widehat{\mathbb{S}}_{\varphi_m^{(c)},\mathcal{G}_m^{(c)}} \cap  \widehat{\mathbb{S}}_{\varphi_w^{(p)},\mathcal{G}_w^{(p)}}=\varnothing$. Fig. \ref{fig:data_geo} illustrates the geometry of signal classes that follow the proposed generative model defined in equations (\ref{eq:genmod_multi}) and (\ref{eq:genmod_scdt}) corresponding to signal and SCDT domains, respectively.

To formulate the solution of the problem defined above, we adapt the subspace-based technique proposed in \cite{rubaiyat2022nearest} for the multiple template-based generative model. First, Let us define a subspace generated by the convex set $\widehat{\mathbb{S}}_{\varphi_m^{(c)},\mathcal{G}_m^{(c)}}$ as: 
\begin{equation}
    \widehat{\mathbb{V}}_m^{(c)} = \text{span}\left(\widehat{\mathbb{S}}_{\varphi_m^{(c)},\mathcal{G}_m^{(c)}}\right).
    \label{eq:subspace}
\end{equation}
Since $\widehat{\mathbb{S}}_{\varphi_m^{(c)},\mathcal{G}_m^{(c)}} \cap \widehat{\mathbb{S}}_{\varphi_w^{(p)},\mathcal{G}_w^{(p)}}=\varnothing$ (when $c\neq p$) and $\left(\mathcal{G}_m^{(c)}\right)^{-1}$ is convex by definition, it is reasonable to assume 
that $\widehat{\mathbb{S}}^{(c)}\cap \widehat{\mathbb{V}}_w^{(p)}=\varnothing $ for any $w=1,...,M_p$ (see Appendix A in supplementary materials for more explanation). Now, if a test sample $s$ is generated according to the generative model defined in eq. (\ref{eq:genmod_multi}), then there exist a certain class `$c$'  and a certain template $\varphi_m^{(c)}$ for which $d^2(\widehat{s}^,\widehat{\mathbb{V}}^{(c)}_m) = 0$. Here, $\widehat{s}$ is the SCDT of the test sample $s$, and $d^2(\widehat{s},\widehat{\mathbb{V}}^{(c)}_m)$ is the Euclidean distance between $\widehat{s}$ and the nearest point in subspace $\widehat{\mathbb{V}}^{(c)}_m$. It also follows, $d^2(\widehat{s},\widehat{\mathbb{V}}^{(p)}_w) > 0$ when $p\neq c$. Therefore, under the assumption that the test sample $s$ is generated according to the generative model for one of the (unknown) classes, the unknown class label can be uniquely predicted by solving,
\begin{equation}
    \argmin_c\min_m~d^2\left(\widehat{s}, \widehat{\mathbb{V}}_m^{(c)}\right), 
    \label{eq:min_problem}
\end{equation}
where $\widehat{\mathbb{V}}_m^{(c)}$ is given by eq. (\ref{eq:subspace}). The proposed algorithm to solve the classification problem is outlined below.

\subsection{Algorithm: nearest local subspace in SCDT domain}

In the signal classification tasks considered below, as usually the case, the template $\varphi^{(c)}_m$ and the deformation set $\mathcal{G}_m^{(c)}$ are usually unknown. Therefore the subset $\widehat{\mathbb{S}}_{\varphi_m^{(c)},\mathcal{G}_m^{(c)}}$ is also unknown; hence, we can not readily estimate $\widehat{\mathbb{V}}_m^{(c)}$ using eq. (\ref{eq:subspace}). Here we devise an algorithm to approximate $\widehat{\mathbb{V}}_m^{(c)}$ using training samples from class $c$. Let us assume that the test sample $s$ is generated according to the generative model $\mathbb{S}_{\varphi_m^{(c)},\mathcal{G}_m^{(c)}}$. 
From eq. (\ref{eq:genmod_scdt}), $\mathbb{S}_{\varphi_m^{(c)},\mathcal{G}_m^{(c)}}$ can be defined in the SCDT domain as:
\begin{align*}
    \widehat{\mathbb{S}}_{\varphi_m^{(c)},\mathcal{G}_m^{(c)}} = \left\{\left(\sum_{i=1}^k\alpha_i f_{i,m}^{(c)}\right)\circ \widehat{\varphi}_m^{(c)}, \alpha_i\geq 0\right\},
\end{align*}
where $\left\{f_{1,m}^{(c)}, f_{2,m}^{(c)},..., f_{k,m}^{(c)}\right\}$ is a set of $k$ linearly independent increasing functions. From this definition, it is evident that $\widehat{\mathbb{V}}_m^{(c)}$ (as defined in eq. (\ref{eq:subspace})) is a $k$ dimensional space. Therefore, if we were to estimate this span, we would need at least $k$ linearly independent elements from the set $\widehat{\mathbb{S}}_{\varphi_m^{(c)},\mathcal{G}_m^{(c)}}$ to model it. 
Since we do not have the knowledge of $\widehat{\mathbb{S}}_{\varphi_m^{(c)},\mathcal{G}_m^{(c)}}$, we employ a nearest local subspace (NLS) search algorithm in SCDT domain to approximate $\widehat{\mathbb{V}}_m^{(c)}$. 
Let us denote the estimated local subspace for class $c$ as $\widetilde{\widehat{\mathbb{V}}}_m^{(c)}$. The solution to the problem defined in eq.(\ref{eq:min_problem}) is then estimated by solving,
\begin{equation}
    \argmin_c ~d^2\left(\widehat{s}, \widetilde{\widehat{\mathbb{V}}}_m^{(c)}\right).
    \label{eq:min_problem_approx0}
\end{equation}
Consider a set of training samples $\left\{s_1^{(c)},..., s_j^{(c)},...,s_{L_c}^{(c)}\right\}\subset \mathbb{S}^{(c)}$ for class $c$, where $L_c$ is the total number of training samples given for class $c$ and $s_j^{(c)}$ is the $j$-th sample. The unknown class of a test sample $s$ is estimated in two steps:

\begin{figure*}[tb]
    \centering
    \includegraphics[width=0.9\textwidth]{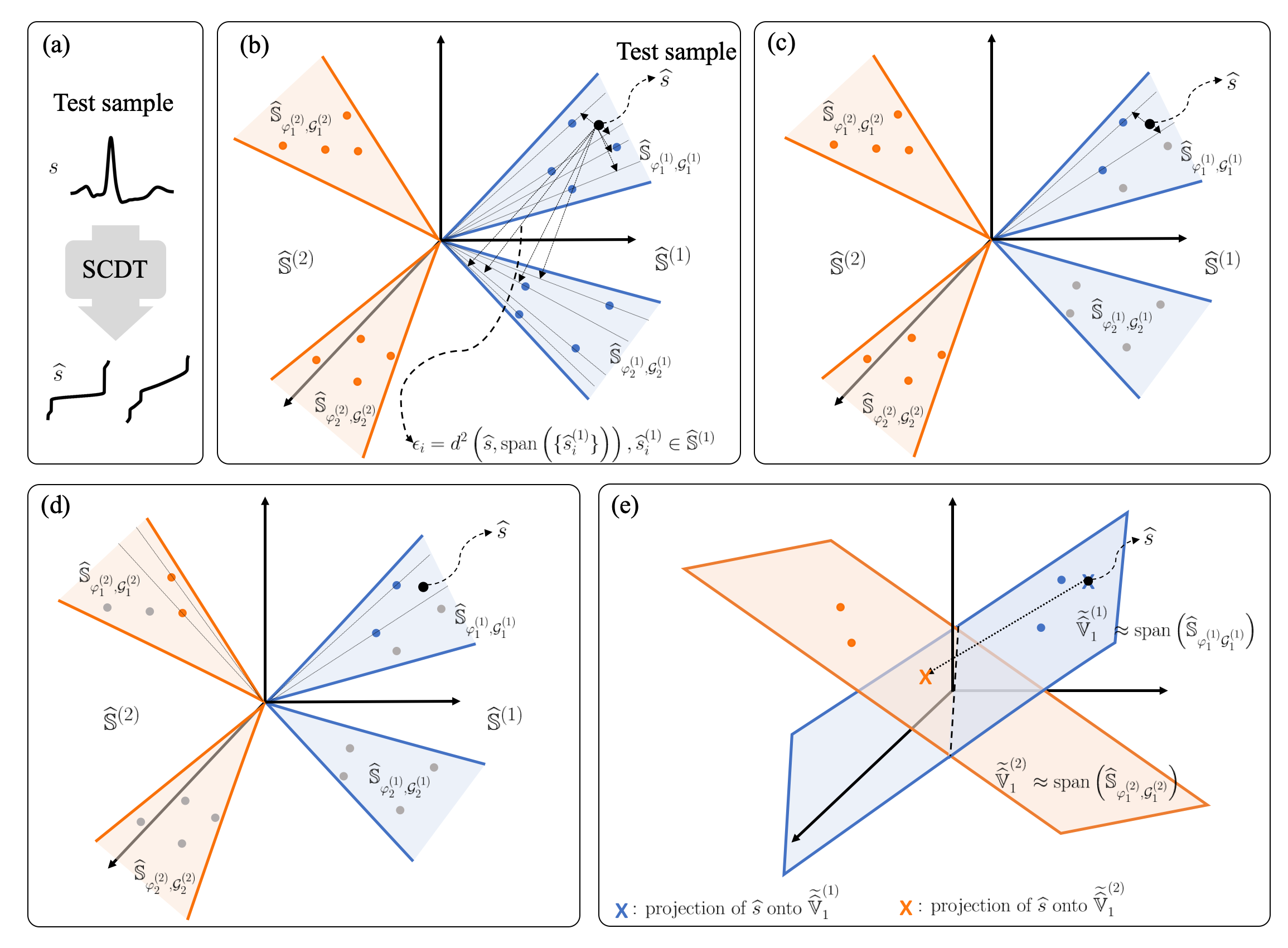}
    \caption{Outline of proposed algorithm: (a) apply SCDT on the test signal $s$ to obtain $\widehat{s}$, (b) measure distance between $\widehat{s}$ and subspace corresponding to each training sample from a particular class (class-1 in this example), (c) find $k$ closest training samples ($k=2$ in this example) to $\widehat{s}$ from class-1, (d) repeat previous steps for other classes (class-2 in this example), (e) build local subspace for each class using the $k$ samples found in previous steps and search for the nearest local subspace to predict the class of $s$.}
    \label{fig:algorithm}
\end{figure*}

\textbf{Step 1:}
We search for the $k$ closest training samples to $\widehat{s}$ from class $c$ based on the distance between $\widehat{s}$ and the span of each training sample. First, we sort the elements from the set $\{\widehat{s}_{1}^{(c)}, ..., \widehat{s}_{L_c}^{(c)}\}$ into $\{\widehat{z}_1^{(c)},...,\widehat{z}_{L_c}^{(c)}\} $ such that 
\begin{align}
	d^2(\widehat{s},\widehat{\mathbb{V}}^{(c)}_{z_1})\leq \cdots \leq d^2(\widehat{s},\widehat{\mathbb{V}}^{(c)}_{z_{l}})\leq \cdots, 
	\label{eq:k-distance}
\end{align}
where $\widehat{\mathbb{V}}^{(c)}_{z_l} = \text{span}\left(\{\widehat{z}_{l}^{(c)}\}\right)$.
Then we pick the first $k$ elements from the sorted set to form $\{\widehat{z}_1^{(c)},\cdots,\widehat{z}_k^{(c)}\}$ for $k\leq L_c$, which gives the set of $k$ closest training samples to $\widehat{s}$ from class $c$ in the above sense (Fig. \ref{fig:algorithm}b-\ref{fig:algorithm}c). We repeat this step for all other classes (Fig. \ref{fig:algorithm}d).

\textbf{Step 2:} Compute $\widetilde{\widehat{\mathbb{V}}}^{(c)}_m$ by:
\begin{align}
    \widetilde{\widehat{\mathbb{V}}}^{(c)}_m = \text{span}\left(\{\widehat{z}_1^{(c)},\cdots,\widehat{z}_k^{(c)}\}\right),
    \label{eq:subspace_knearest}
\end{align}
which approximates the nearest local subspace from class $c$ with respect to $\widehat{s}$. We then predict the unknown class of the test sample $s$ by solving eq. (\ref{eq:min_problem_approx0}). 
 Fig. \ref{fig:algorithm}e illustrates the second step. Note that similar local subspace classification techniques can be found in the literature \cite{laaksonen1997local} \cite{cevikalp2007local}. Here, we employ the nearest local subspace search technique in the SCDT domain to exploit the properties of the SCDT that simplify the classification problem described above.
Next, we show that the performance of the classifier can further be improved by utilizing the composition property of the SCDT through analytical enrichment of the subspace method.

\subsubsection{Subspace enrichment}
The algorithm outlined above searches for the $k$ closest training samples to a given test sample in SCDT domain, and then forms a local subspace using these $k$ samples. Inspired by \cite{shifat2020radon,rabbi2022invariance}, here we adapt the technique to mathematically enrich the ensuing subspace with certain prescribed deformations. Here we enrich $\widetilde{\widehat{\mathbb{V}}}^{(c)}_m$ in such a way that it will automatically include the samples undergoing certain time deformations. The spanning sets corresponding to certain specific deformations are derived below:
\begin{itemize}
    \item Translation: In case of translation, $g(t)$ is given by $t-\mu$, where $\mu \in \mathbb{R}$ is the translation parameter. Using the composition property of the SCDT, the transform of the translated signal $s_g(t) = s(t - \mu)$ is given by $ \widehat{s}_g=\left((s^{+})^*+\mu,\|s^{+}\|_{L_1},  (s^{-})^*+\mu,\|s^{-}\|_{L_1}\right)$. It implies that the translation applied to the signal along the independent axis results in translation along the dependent axis in SCDT domain. Hence, the spanning set for translation is defined as $\mathbb{U}_T=\{u(t)\}$, where $u(t)=1$.
    \item Dilation: A time-dilated (scaled) version of a signal $s(t)$ is defined as: $s_g(t)=\alpha s(\alpha t), \alpha\in\mathbb{R}^+$. The transform of the signal $s_g(t)$ is given by: $\widehat{s}_g = \left( \frac{(s^{+})^*}{\alpha},\|s^{+}\|_{L_1}, \frac{(s^{-})^*}{\alpha},\|s^{-}\|_{L_1}, \right)$. An additional spanning set is not required for dilation, as it is inherent in the modeling of the subspace.
    \item Time-warpings other than translation and dilation are also observed in certain classification problems. To include those deformations in the SCDT domain, we approximate the increasing function $g^{-1}\circ \widehat{z}_l^{(c)}$ as:
    \begin{align}
        g^{-1}\circ \widehat{z}_l^{(c)} = \begin{cases}
        \sum\limits_{n=-N}^N c_n \zeta_n(\widehat{z}_l^{(c)}) &\text{for $n\neq 0$}\\
        c_0 \widehat{z}_l^{(c)} &\text{for $n=0$}
        \end{cases}
        \label{eq:approx_ginv}
    \end{align}
    where, $c_n>0, \quad\sum_{n=-N}^N c_n = 1$, and
    \begin{align}
        \zeta_n(\widehat{z}_l^{(c)}) = \left[\widehat{z}_l^{(c)} - \frac{\sin{n\pi\widehat{z}_l^{(c)}}}{|n|\pi}\right].\nonumber
    \end{align}
    For the set $\{\widehat{z}_1^{(c)},\cdots,\widehat{z}_k^{(c)}\}$ from eq. (\ref{eq:subspace_knearest}), the spanning set is given by the set $\mathbb{U}_H = \{\zeta_n(\widehat{z}_1^{(c)}),...,\zeta_n(\widehat{z}_k^{(c)})\}$, for $n=-N,...,-1,1,...,N$.
\end{itemize}
In light of the discussion above, the subspace $\widetilde{\widehat{\mathbb{V}}}^{(c)}_m$ in eq. (\ref{eq:subspace_knearest}) can be enriched as follows:
\begin{align}
    \widetilde{\widehat{\mathbb{V}}}^{(c)}_{m} = \text{span}\left(\{\widehat{z}_1^{(c)},\cdots,\widehat{z}_k^{(c)}\}\cup \mathbb{U}_T \cup \mathbb{U}_H\right).
    \label{eq:enrichedV}
\end{align}
Note that the subspace $\widehat{\mathbb{V}}^{(c)}_{z_l}$ in (\ref{eq:k-distance}) can also be enriched in similar manner, i.e. $\widehat{\mathbb{V}}^{(c)}_{z_l} = \text{span}\left(\{\widehat{z}^{(c)}_l\}\cup \mathbb{U}_T\cup \mathbb{U}_H \right)$, where $\mathbb{U}_H = \{\zeta_n(\widehat{z}_l^{(c)})\}$, for $n=-N,...,-1,1,...,N$.

\subsubsection{Training phase} 
In the training phase of the algorithm, the subspace corresponding to each of training sample is calculated. The first step is to compute SCDTs for all training samples from class $c$. Then, we take a training sample $\widehat{s}_{l}^{(c)}$ and orthogonalize $\{\widehat{s}_{l}^{(c)}\}\cup \mathbb{U}_T\cup \mathbb{U}_H$ to obtain the basis vectors that span the enriched subspace corresponding to that sample. Let $B_{l}^{(c)} = \left[b_{l,1}^{(c)}, b_{l,2}^{(c)}, ... \right]$ be a matrix that contains the basis vectors in its column. We repeat these calculations for all the training samples to form $B_{l}^{(c)}$ for $l=1,2,...,L_c$ and $c=1,2,...,$ etc.

\subsubsection{Testing phase}
The testing algorithm begins with taking SCDT of the test sample $s$ to obtain $\widehat{s}$ followed by the nearest local subspace search in SCDT domain. In the first step of the algorithm, we estimate the distance of the subspace corresponding to each of the training samples from $\widehat{s}$ by:
\begin{align*}
    \epsilon_l = \|\widehat{s} - B_{l}^{(c)}B_{l}^{(c)^T}\widehat{s}\|^2,\qquad l=1,2,...,L_c,
\end{align*}
where $\|.\|$ denotes $L_2$ norm. Note that $B_{l}^{(c)}B_{l}^{(c)^T}$ is
the orthogonal projection matrix onto the space generated by
the span of the columns of $B_{l}^{(c)}$ (computed in the training phase). 
We then find $\{\widehat{z}_1^{(c)},\cdots,\widehat{z}_k^{(c)}\}$, a set of $k$ closest training samples to the test sample $\widehat{s}$ from class $c$, based on the distances ${\epsilon_1,...,\epsilon_{L_c}}$.
In the next step, we orthogonalize $\{\widehat{z}_1^{(c)},\cdots,\widehat{z}_k^{(c)}\}\cup \mathbb{U}_T\cup \mathbb{U}_H$ to obtain the basis vectors $\{b_1^{(c)},b_2^{(c)},...\}$ spanning the local subspace from class $c$ with respect to $\widehat{s}$. Let $B^{(c)}=\left[b_1^{(c)},b_2^{(c)},... \right]$ for $c=1,2,...,$ etc. The unknown class of $s$ is then estimated by:
\begin{equation}
    \arg\min_c~ \|\widehat{s} - B^{(c)}B^{(c)^T}\widehat{s}\|.
    \label{eq:test_step}
\end{equation}

Note that the proposed algorithm requires two parameters $k$ and $N$ (see eq. (\ref{eq:approx_ginv})) to be tuned prior to the testing phase. We use a validation set split from the training set to estimate the optimum values for these parameters. We then follow the steps of the proposed algorithm outlined above with the validation set for varying $k$ and $N$. Parameter values corresponding to the best validation accuracy are chosen to be used in the testing phase. This step is done during the training phase.

\subsection{Proof-of-concept simulation}
\begin{figure}[tb]
    \centering
    \includegraphics[width=0.47\textwidth]{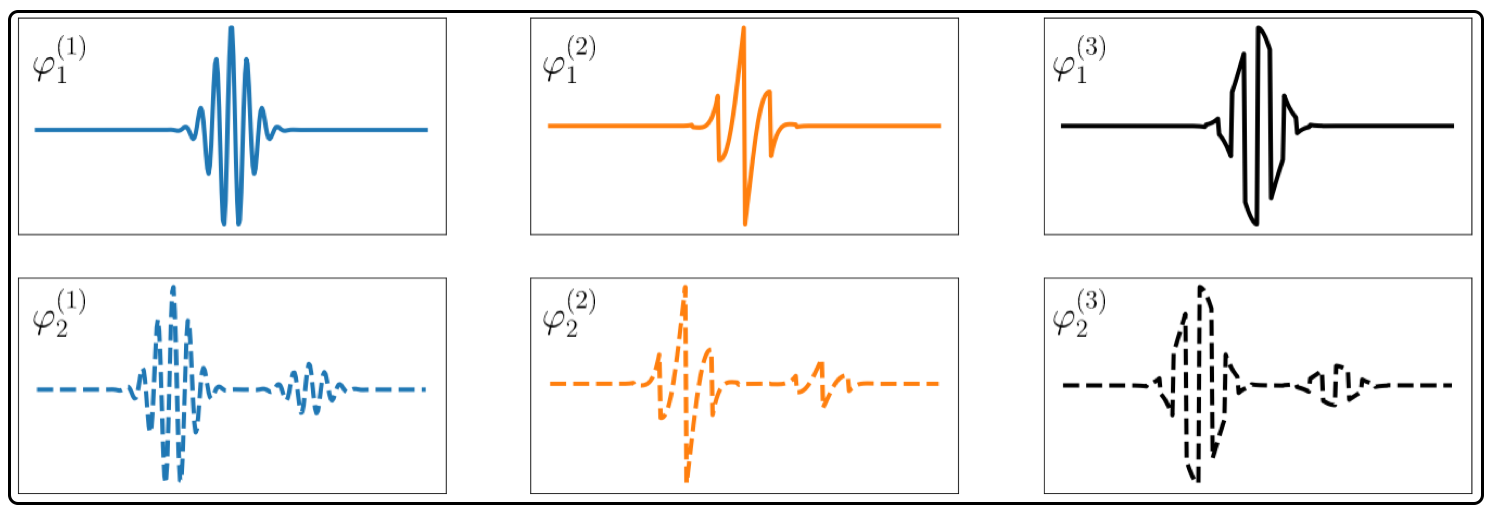}
    \includegraphics[width=0.45\textwidth]{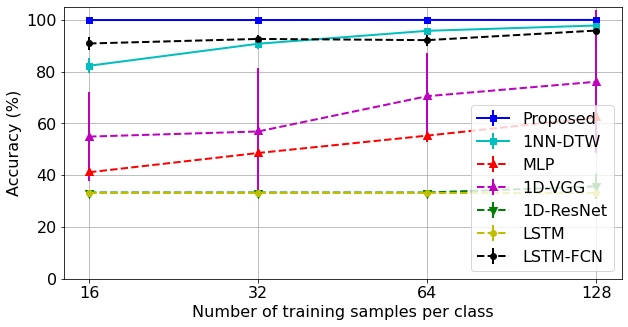}
    \caption{A simulated experiment to demonstrate the efficacy of the proposed method in classifying signal classes that follow the generative model defined in eq. (\ref{eq:genmod_multi}).}
    \label{fig:res_synthetic}
    \vspace{-1em}
\end{figure}

To demonstrate the efficacy of the proposed algorithm in solving the classification problem stated in section \ref{sec:problem} we performed a simulated experiment. We took six prototype signals shown in Fig. \ref{fig:res_synthetic} as the templates ($\varphi_{m}^{(c)}$) corresponding to three different classes ($c=1,2,3$), i.e., each class has two templates ($m=1,2$). We then generated a synthetic dataset by applying specific time deformations on the prototype signals as follows:
\begin{align}
    &s^{(c)}_j = g'_j(t)\varphi^{(c)}_m(g_j(t)), \nonumber\\
    &g_j(t) = \omega\zeta(t)+\tau,\quad g'_j(t)>0, \nonumber\\
    &\zeta'(t) = \frac{\partial}{\partial t}\zeta(t) = \sum_{n=1}^N\alpha_n \frac{1}{\sqrt{2\pi}w_n} e^{-\frac{1}{2}\left(\frac{t-\mu_n}{w_n}\right)^2}
    \label{eq:g_synthetic}
\end{align}
where $\alpha_n>0,\quad \sum_n \alpha_n=1$ and the parameter values used to calculate $g_j(t)$ are randomly chosen from fixed intervals. The dataset was equally split into training and testing sets. The proposed classification method was then trained with a varying number of training samples per class randomly chosen from the training set and evaluated on the testing set. Note that the samples were chosen in such a way that the training set contains an equal number of samples generated from each template. 
Fig. \ref{fig:res_synthetic} shows the test accuracy plots with respect to the number of training samples per class. It demonstrates that the proposed method achieves the perfect classification accuracy with few training samples (obtained $99.97\%$ test accuracy with only $16$ training samples per class), while some alternative end-to-end systems (discussed in next section) fail to achieve such performance even with higher number of training data. That means, if the signal classes follow the generative model defined in eq. (\ref{eq:genmod_multi}), the proposed method solves the signal classification problem stated in section \ref{sec:problem}.

\section{Experiments and Results} \label{sec:exp}

\subsection{Experimental setup}
The goal is to evaluate the performance of the proposed generic end-to-end classifier with respect to selected state-of-the-art end-to-end time series classification techniques. The classification performance of the different methods were studied in terms of test accuracy, data efficiency, computational efficiency, and robustness to the out-of-distribution samples. We conducted experiments on several time series data and compared the results against several methods: Multilayer Perceptrons (MLP) \cite{iwana2021empirical}, 1D Visual Geometry Group (VGG) \cite{iwana2021empirical}, 1D Residual Network (ResNet) \cite{fawaz2018data} \cite{iwana2021empirical}, Long Short Term Memory (LSTM) \cite{hochreiter1997long} \cite{iwana2021empirical}, Long Short Term Memory Fully Convolutional Network (LSTM-FCN) \cite{karim2017lstm} \cite{iwana2021empirical}, and 1-nearest neighbor DTW (1NN-DTW) \cite{xi2006fast}. For the neural network methods, we used the implementations outlined in \cite{iwana2021empirical} without any data augmentation. During the training process, $10\%$ of the training samples were used as the validation set, and the test performance was reported based on the model that had the best validation performance. In case of the 1NN-DTW method, the validation set was used to tune the parameter corresponding to the warping window size. To show the efficacy of the proposed multiple template-based generative model (defined in eq. (\ref{eq:genmod_multi})) over the single template-based model (eq. (\ref{eq:genmod_single})), we also compared the results against the SCDT-NS classifier \cite{rubaiyat2022nearest} proposed earlier.

We followed the training and testing procedures outlined in the previous section for the proposed method. The orthogonalization operations were performed using singular value decomposition (SVD). The left singular vectors obtained by the SVDs were used to construct the matrices $B^{(c)}_l$ and $B^{(c)}$. Following \cite{shifat2020radon}, the number of the basis vectors was chosen in such a way that the sum of variances explained by the selected basis vectors captures at least $99\%$ of the total variance explained by all the samples in the sorted set $\{\widehat{z}_1^{(c)},\cdots,\widehat{z}_k^{(c)}\}$. The SCDTs were computed with respect to a 1D uniform probability density function.

\subsection{Datasets} \label{sec:datasets}
\begin{figure*}[tb]
    \centering
    \includegraphics[width=0.9\textwidth]{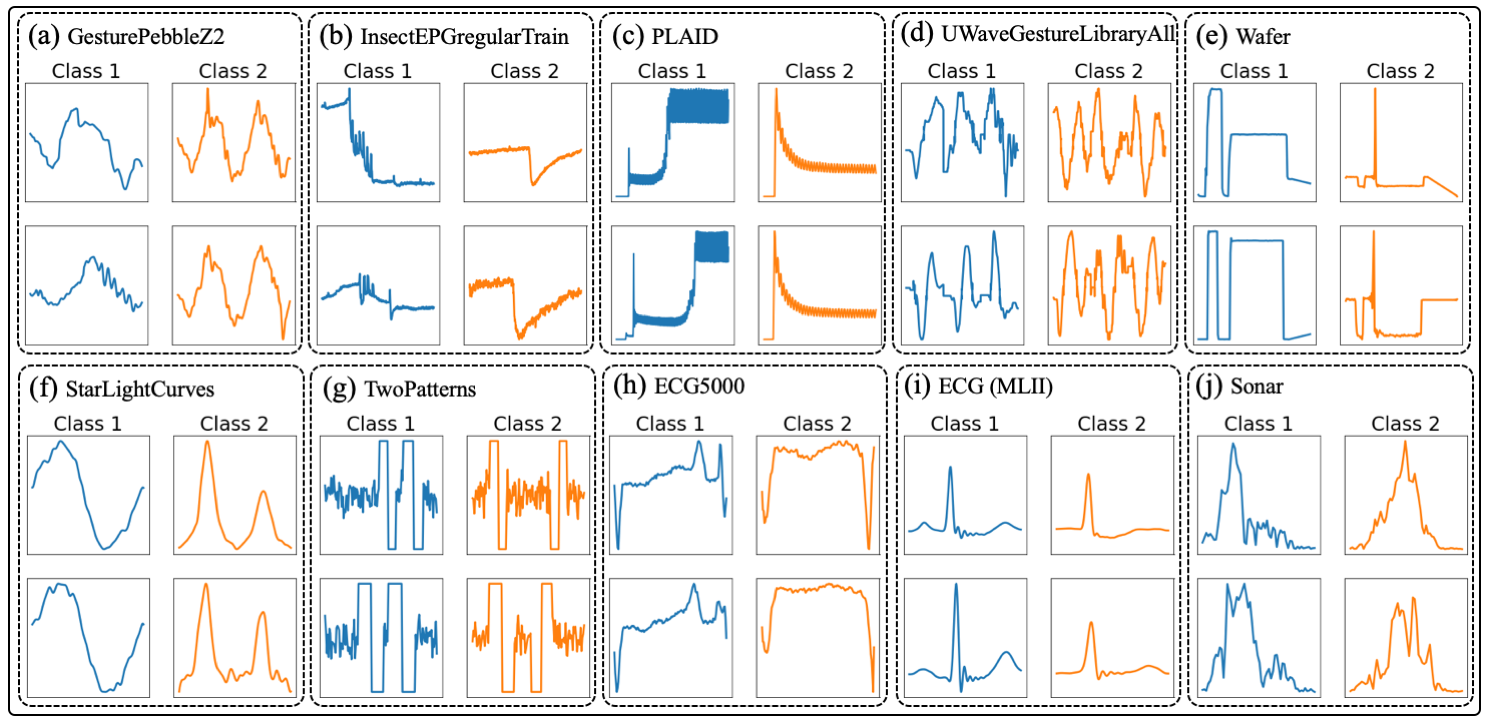}
    \caption{Some example signals from the datasets used to evaluate the proposed method. 
    }
    \label{fig:sig_dataset}
\end{figure*}

To evaluate the comparative performance of the proposed method with respect to other end-to-end classifiers we selected multiple datasets with signal classes representing well-defined time series events. For example, the accelerometer data plotted in Fig. \ref{fig:sig_dataset}(a) represent particular hand gestures. Similarly, signals shown in Fig. \ref{fig:sig_dataset}(i) represent either normal or abnormal heartbeats.
With a focus towards this condition, we identified 10 different time series datasets, 8 of which were downloaded from the UCR time series classification archive \cite{UCRArchive2018}. 
Some example signals from these datasets are shown in Fig. \ref{fig:sig_dataset}. 
Details about the datasets are given below:

\begin{itemize}
    \item \textit{GesturePebbleZ2} \cite{mezari2017gesture}: Accelerometer data collected using Pebble smart watch from 4 different persons performing 6 hand gestures.  (classes: 6, train samples: $22\sim 25$ per class, test samples: $25\sim 32$ per class).
    \item \textit{InsectEPGRegularTrain} \cite{willett2016machine}: Contains electrical penetration graph (EPG) data which capture voltage changes of the electrical circuit that connects insects and their food source. (classes: 2, train samples: $22\sim 30$ per class, test samples: $89\sim 118$ per class).
    \item \textit{PLAID}: Plug Load Appliance Identification Dataset \cite{gao2014plaid}. The data are intended for load identification research using transient voltage/current measurements from 11 different appliance types. 
    (classes: 11, train samples: $13\sim 88$ per class, test samples: $13\sim 87$ per class).
    \item \textit{UWaveGestureLibraryAll} \cite{liu2009uwave}: A set of eight simple gestures generated from accelerometers using Wii remote. (classes: 8, train samples: $100\sim 127$ per class, test samples: $433\sim 460$ per class).
    \item \textit{Wafer} \cite{olszewski2001generalized}: A collection of inline process control measurements recorded from various sensors during the processing of silicon wafers. The two classes are normal and abnormal, with a significant class imbalance. Hence, a subset of the original dataset is used to ensure the class balance. (classes: 2, train samples: $97\sim 100$ per class, test samples: $665\sim 700$ per class).
    \item \textit{StarLightCurves} \cite{rebbapragada2009finding}: Collection of time series signals representing the brightness of celestial objects as a function of time. (classes: 3, train samples: $150$ per class, test samples: $500$ per class).
    \item \textit{TwoPatterns} \cite{geurts2002contributions}: A simulated dataset (classes: 4, train samples: $237\sim 271$ per class, test samples: $959\sim 1035$ per class).
    \item \textit{ECG5000} \cite{UCRArchive2018}: A subset of BIDMC Congestive Heart Failure Database (CHFDB) downloaded from PhysioNet. With a purpose of evaluating the methods trained with a large set, we interchanged the train and test sets of the original dataset. (classes: 2, train samples: $1873\sim 2627$ per class, test samples: $208 \sim 292$ per class).
    \item \textit{ECG (MLII)} \cite{ecg744}: A subset of a publicly available dataset reported in \cite{plawiak2018novel}. The ECG signals were collected from the PhysioNet MIT-BIH Arrythmia database \cite{mousavi2019inter}. A method described in \cite{bassiouni2018intelligent} was used to segment the heartbeats from the ECG fragments. (classes: 3, train samples: $200$ per class, test samples: $200$ per class).
    \item \textit{Connectionist Bench (Sonar, Mines vs. Rocks)} \cite{Dua:2019} \cite{gorman1988analysis}: This dataset contains energy patterns (with respect to frequency bands) of the signals obtained by bouncing sonar signals off a metal cylinder and some rocks at various angles and under various conditions. 
    (classes: 2, train samples: $49\sim 55$ per class, test samples: $48\sim 56$).
\end{itemize}
In addition to the datasets listed above, we have used another dataset where time series events are not well-defined as an example where the proposed method is not expected to work well. We have used the gearbox fault diagnosis data \cite{gearbox_openEI}, a collection of vibration signals recorded by using SpectraQuest's Gearbox Fault Diagnostics Simulator. The dataset was recorded in two different scenarios: 1) healthy and 2) broken tooth conditions.

\subsection{Test accuracy}

\begin{table*}[h!]
    \centering
    \resizebox{1.8\columnwidth}{!}{
    \begin{tabular}{|l|c|c|c|c|c|c|c|c|}
    \hline
    Dataset & MLP & 1D-VGG & 1D-ResNet & LSTM & LSTM-FCN & SCDT-NS & 1NN-DTW &  \makecell[c]{\textbf{Proposed}} \\
    \hline
    GesturePebbleZ2 & 83.92 & 48.99 & 63.23 & 60.19 & 85.82 & 81.64 & \textbf{94.94} & 94.30\\
    InsectEPGRegularTrain & 80.92 & 82.07 & 50.58 & 57.05 & 59.95 & 73.43 & 82.61 & \textbf{90.82}\\
    PLAID & 39.65 & 19.35 & 29.81 & 26.35 & 39.57 & 58.29 & \textbf{73.56} & 70.95\\
    UWaveGestureLibraryAll & 94.74 & \textbf{96.42} & 42.86 & 30.72 & 94.16 & 90.4 & 94.22 & 94.95\\
    Wafer & 68.69 & 94.03 & 53.66 & 51.36 & 86.70 & 92.75 & \textbf{98.16} & 95.60\\
    StarLightCurves & 75.85 & 70.0 & 56.53 & 60.54 & 74.38 & 77.4 & 83.80 & \textbf{84.2}\\
    TwoPatterns & 88.12 & 95.15 & 87.42 & 80.62 & 98.47 & 95.15 & 99.90 & \textbf{99.92}\\
    ECG5000 & 98.9 & 98.92 & 98.92 & 98.36 & 98.76 & 93.4 & \textbf{99.2} & 97.6\\
    ECG (MLII) & 33.33 & 33.33 & 56.55 & 44.20 & 66.42 & 56.67 & 45.67 & \textbf{68.83}\\
    Sonar & 68.36 & \textbf{80.57} & 53.85 & 53.85 & 53.85 & 61.54 & 76.92 & 78.85 \\
    \hline
    Win  & 0 & 2 & 0 & 0 & 0 & 0 & \textbf{4} & \textbf{4} \\
    AVG arithmetic ranking &  4.6 & 4.3 & 6.1 & 7.0 & 4.7 & 4.7 & \textbf{2.2} & \textbf{2.2}  \\
    AVG geometric ranking & 4.46 & 3.38 & 5.73 & 6.94 & 4.4 & 4.48 & 1.85 & \textbf{1.84}  \\
    MPCE  & 0.084 & 0.072 & 0.123 & 0.128 & 0.079 & 0.071 & 0.049 & \textbf{0.038} \\
    \hline
    \end{tabular}
    }
    \caption{Test accuracy ($\%$), rank-based statistics, and MPCE calculated for the classifiers across different datasets.}
    \label{table:test_acc}
    \vspace{-1em}
\end{table*}

To demonstrate the efficacy of the proposed method as a generic classifier, we applied it to the datasets listed above and compared the test accuracies against the aforementioned end-to-end classification methods. Table \ref{table:test_acc} shows the results and a comprehensive comparison with five neural network-based classifiers, 1NN-DTW, and SCDT-NS. 
The results reported in the table show that in most of the datasets (8 out of 10), the proposed method outperformed the deep learning methods and provided competitive test accuracy with respect to 1NN-DTW. The average arithmetic and geometric rankings also demonstrate the efficacy of the proposed method as a generic end-to-end technique to classify segmentable time series events.

Besides test accuracy and rank-based statistics, we also calculated mean per class error (MPCE) for each method. This metric was proposed in \cite{wang2017time} to evaluate the performance of a generic classifier on multiple datasets. MPCE is defined as the arithmetic mean of the per class error (PCE) which is calculated for $i$-th model on $j$-th dataset as: $\text{PCE}_{i,j} = \frac{e_{i,j}}{c_{j}}$. Here $e_{i,j}$ is the error rate of $i$-th model on $j$-th dataset, and $c_{j}$ is the total number of classes present in $j$-th dataset. MPCE for the corresponding model is given by,
\begin{equation*}
    \text{MPCE}_i = \frac{1}{J}\sum_{j=1}^J \text{PCE}_{i,j},
\end{equation*}
where $J$ is the total number of datasets used in the experiment. The MPCE values reported in Table \ref{table:test_acc} indicate that the proposed method generates the least expected error rate per class across all the datasets in comparison to other classifiers.

\subsection{Data efficiency}\label{sec:data_efficient}
\begin{figure}
     \centering
    \includegraphics[width=0.45\textwidth]{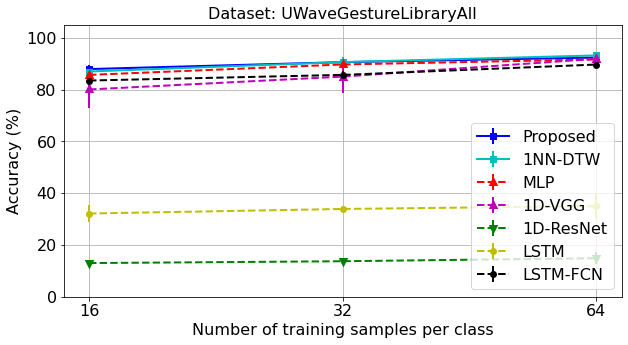}
    \includegraphics[width=0.45\textwidth]{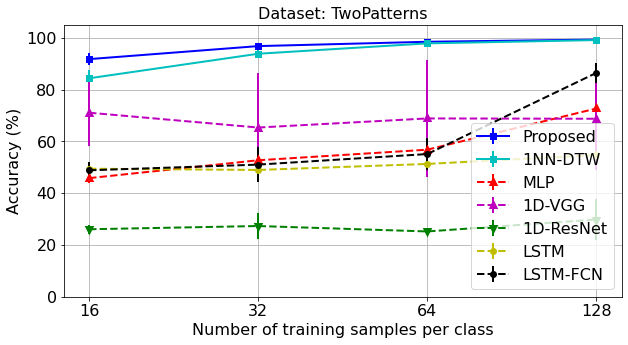}
    \caption{Accuracy as a function of number of training samples per class for different classification methods.}
    \label{fig:accVStrsamples}
    \vspace{-1em}
\end{figure}

To show the data efficiency of the proposed method, we set up an experiment where we trained the models with a varying number of training samples per class. For a training split of a particular size, its samples were randomly drawn from the original training set, and the experiments for this particular size were repeated 10 times. Fig. \ref{fig:accVStrsamples} shows average accuracies with respect to the number of training samples per class for two datasets: \textit{UWaveGestureLibraryAll} and \textit{TwoPatterns}. The standard deviation for each split is also shown using the error bar. The plots illustrate that the proposed method achieves higher accuracy than the deep learning methods with fewer training samples. Similar results can be seen in other datasets as well (see Appendix B in supplementary materials). 

\subsection{Computational efficiency}
\begin{figure}
     \centering
    \includegraphics[width=0.45\textwidth]{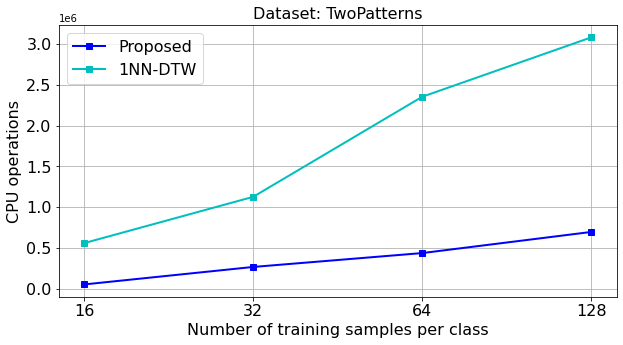}
    \caption{Computational complexity plots of 1NN-DTW and the proposed method as a function of number of training samples. Complexities of these two methods are given by $O(pnr)$ and $O(ck^2n)$, respectively, where $c$ is the number of classes, $p$ is the total number of training samples, $n$ is the signal length, $r$ is the window length to calculate DTW, and $k$ is the number of closest samples with respect to a test sample used to build the local subspace in proposed method.}
    \label{fig:comp_plots}
\end{figure}

\begin{figure}
     \centering
    \includegraphics[width=0.45\textwidth]{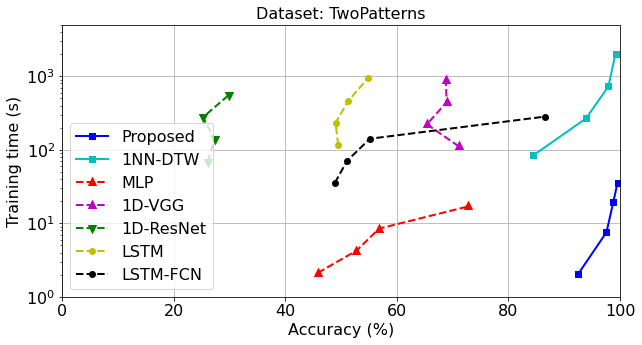}
    \includegraphics[width=0.45\textwidth]{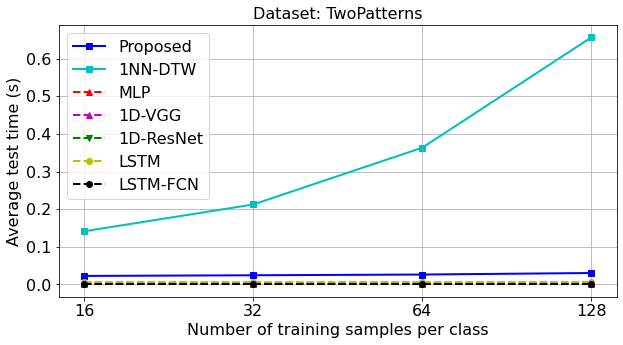}
    \caption{(Upper) Training time in seconds vs test accuracy for different classification methods. (Lower) Average time taken by the classification methods to predict the class of a test signal. Experiments were run using Python version 3.6.9 on a computer with an Intel(R) Xeon(R) CPU E5-2630 v4 processor running at 2.20GHz using 62 GB of RAM.}
    \label{fig:time_tr_te}
\end{figure}

The proposed classification technique is not only effective in classifying time series events but also computationally very efficient. Fig. \ref{fig:comp_plots} shows the computational complexity plots of the proposed method and 1NN-DTW as a function of number of training samples per class for \textit{TwoPatterns} dataset. It demonstrates that the proposed classifier requires less CPU operations in comparison to 1NN-DTW. Fig. \ref{fig:time_tr_te} (upper panel) plots the average time (in seconds) required during the training phase as a function of accuracy for each method. It illustrates that all the alternative end-to-end solutions require greater number of computations to train the models with respect to the proposed solution to achieve same level of accuracy. Despite the fact that our method searches for $k$-closest training samples for a given test signal, it still provides competitive performance as shown in Fig. \ref{fig:time_tr_te} (lower panel) in terms of test time in comparison to deep learning-based methods. The plots of the average time (in seconds) taken by the classification methods to test a signal from \textit{TwoPatterns} dataset show that the test time of the proposed method is close to the neural network-based classifiers, while 1NN-DTW is highly expensive in terms of computation during testing phase. Note that the SCDT is currently implemented with \textit{for} loops in python, which is less than ideal in terms of execution time. A compiled language (e.g. C, C++) would execute \textit{for} loops much faster.

\subsection{Robust to out-of-distribution samples}
\begin{figure}[tb]
\centering
\includegraphics[width=0.45\textwidth]{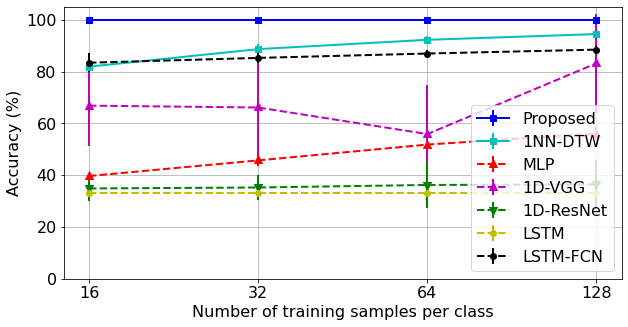}
\caption{Accuracy as a function of number of training samples per class under the out-of-distribution setup.}
\label{fig:accVStrOutDist}
\vspace{-1.5em}
\end{figure}

To demonstrate the robustness to out-of-distribution examples, we adopted a similar concept used in \cite{rubaiyat2022nearest}. We generated a synthetic dataset by applying time deformations defined in eq. (\ref{eq:g_synthetic}) on three prototype signals: a Gabor wave, an apodized sawtooth wave, and an apodized square wave (shown in the top row of Fig. \ref{fig:accVStrsamples}). Note that we used a single template per class to generate data for maintaining a simple experimental setup. We varied the magnitude of the confounding factors (i.e., the parameters used to calculate $g_j(t)$) to generate different distributions for training and testing sets. The `in-distribution’ set used during the training process consisted of signals with parameter values chosen randomly from smaller intervals with respect to the `out-distribution' (testing) set. Table \ref{tab:outdistparams} shows the list of the parameters of interest for the `out-of-distribution' experiment. The intervals (from which the parameter values were chosen) corresponding to the training and testing sets are also given in the table. Fig. \ref{fig:accVStrOutDist} shows test accuracies with respect to the number of training samples per class for the comparing methods. The plot shows that the proposed method significantly outperforms other end-to-end solutions with very few training samples. Meaning, the proposed classification technique provides the best classification result if the test signal belongs to the `out-distribution' set in the above sense but follows the generative model discussed in the previous section.

\begin{table}[!htb]
    \centering
    \normalsize
    \begin{tabular}{lll}
    \hline
    Parameter                & Train  & Test    \\ \hline
    $N$ & $\left[2,5\right]$ & $\left[2,10\right]$ \\
    $\mu_n$ & $\mathcal{N}(0.5,0.2)$ & $\mathcal{N}(0.5,0.3)$ \\
    $\omega$ & $\left[0.9,1.1\right]$ & $\left[0.75,1.25\right]$ \\
    $\tau$ & $\left[-0.05,0.05\right]$ & $\left[-0.1,0.1\right]$ \\
    \hline
    \end{tabular}
    \caption{Intervals used in the out-of-distribution setup.}
    \label{tab:outdistparams}
    \vspace{-1em}
\end{table}

\subsection{A dataset that does not follow generative model}
\begin{figure}[tb]
    \begin{subfigure}{.47\textwidth}
        \includegraphics[width=\textwidth]{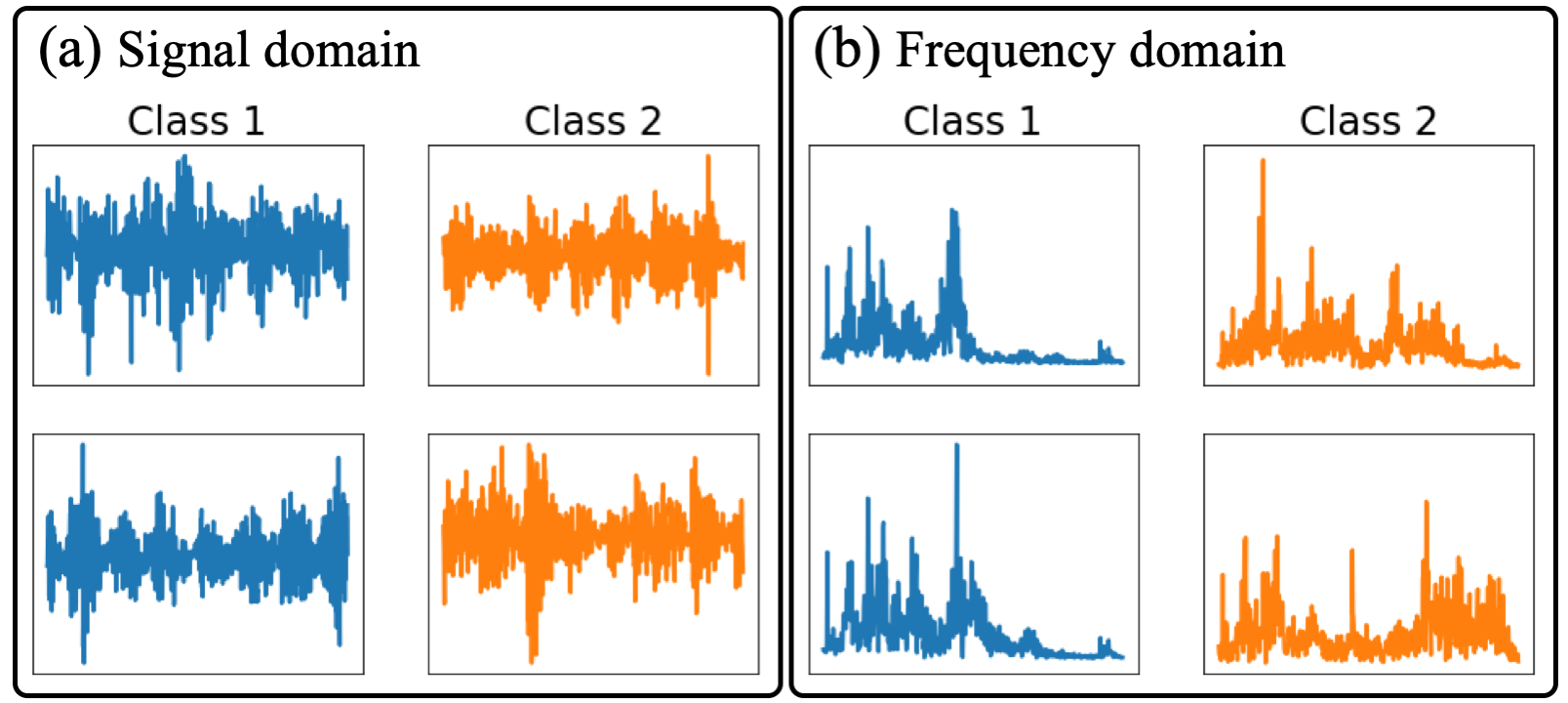}
    \end{subfigure}
    \begin{subfigure}{0.45\textwidth}
        \vspace{1em}
        \centering
        \begin{tabular}{|l|c|c|}
        \hline
            Method & Time domain & Frequency domain \\
            \hline
            MLP & 55.67 & 90.45 \\
            1D-VGG & 50.77 & 56.02 \\
            1D-ResNet & 71.5 & 50.0 \\
            LSTM & 57.32 & 72.35 \\
            LSTM-FCN & \textbf{76.5} & 90.92 \\
            SCDT-NS & 52.75 & 91.5 \\
            1NN-DTW & 75.75 & \textbf{93.5} \\
            \textbf{Proposed} & 57.25 & 89.5 \\
            \hline
        \end{tabular}
    \end{subfigure}
    
    \caption{Plots in (a) show few examples from the gearbox fault diagnosis dataset that do not follow the generative model. Plots in (b) show the signals in frequency domain corresponding to the signals shown in (a), which seem to fit the generative model; hence, the proposed method performs better in classifying these signals. Table in lower panel shows the test accuracy ($\%$) generated by the classifiers on both time and frequency domain data.
    }
    \label{fig:data_gearbox}
    \vspace{-1em}
\end{figure}

As described in \ref{sec:datasets}, the proposed method performs well if the time series data is well segmented, i.e. contains events with well-defined start and end points. There are examples of signal classification problems where data do not follow these conditions. One such example is the gearbox fault diagnosis dataset \cite{gearbox_openEI}, where vibration signals are used to detect a faulty gearbox. Fig. \ref{fig:data_gearbox}(a) shows that the vibration signals do not possess segmentable time series events. Hence, the proposed method is not suitable for classifying raw gearbox vibration data, as shown in the table in Fig. \ref{fig:data_gearbox}. However, we can transform each time series to frequency domain using the Fourier transform \cite{bracewell1986fourier}, where signals of finite length are necessarily band limited (and thus have a beginning and end in frequency domain). As expected the proposed method performs better when applied to the frequency domain signals shown in Fig. \ref{fig:data_gearbox}(b).

\section{Discussion}\label{sec:diss}
Classification test accuracies, rank-based statistics, and MPCE reported in Table \ref{table:test_acc} across 10 different time series datasets suggest that the proposed method is a very good generic end-to-end signal classification model for time series containing segmented events. Moreover, Fig. \ref{fig:accVStrsamples} shows that the proposed method can achieve high accuracy with few training samples. The computational efficiency of the proposed classifier is also demonstrated in figures \ref{fig:comp_plots} and \ref{fig:time_tr_te} in terms of CPU operations, and average training and testing time with respect to other end-to-end classifiers. It is also evident from Table \ref{table:test_acc} that the proposed method outperforms the SCDT-NS classifier \cite{rubaiyat2022nearest} which uses a single template-based generative model.

Another compelling property of the proposed method is the robustness to out-of-distribution samples since it generalizes to samples outside the known distribution when the signal classes conform to the specific generative model. Plots in Fig. \ref{fig:accVStrOutDist} show that other methods fail to achieve good performance under the out-of-distribution setup, whereas the proposed method achieves perfect test accuracy with very few training samples ($\sim 16$ per class). The reason behind the robustness to the out-of-distribution setup is that the proposed method is capable of learning the underlying data model, more specifically, the types of unknown deformations present in the signals. It can then successfully classify an unknown signal in presence of such deformations but with different magnitudes. 

The main assumption of the proposed method is that the dataset needs to conform to the underlying generative model proposed earlier. Specifically, we showed above that the method works best when the time series (signal) being classified contains segmented events with finite duration. Fig. \ref{fig:data_gearbox} shows an example where raw signals (from a gearbox vibration experimental setup) do not possess well-defined time series events of finite duration. Hence, the proposed method performs poorly in classifying those signals. However, the same signals in frequency domain seem to fit the generative model better and the proposed method performs better in classifying the gearbox data in frequency domain.

To summarize, this paper presents a new idea of representing segmented signal data using a generative model such that signals from a particular class can be considered as observations of a set of unknown templates under some unknown deformations. Under this assumption, we formulated a classification problem for segmented signal classes. Then we showed that if the data follow the proposed generative model, a simple solution can be devised by searching nearest local subspace in SCDT domain. Through extensive experiments, we demonstrated that the proposed solution is effective in classifying unknown signals, computationally very cheap, data efficient, and robust to out-of-distribution samples.

\section{Conclusion}\label{sec:conc}
This paper introduced a new end-to-end signal classification method based on a recently developed signal transform. First, we formulated the problem statement based on a multiple template-based generative model observed under unknown deformations. Then, we proposed an end-to-end solution to the problem by employing a nearest local subspace search algorithm in SCDT domain. Although the problem statement and solution are based on the assumption that signals are observations of templates under confounding deformations, knowledge of these templates or confound deformations is not required.  The model was demonstrated to achieve high test accuracy across multiple time series datasets. Several experiments show that the proposed method not only outperforms the state-of-the-art deep learning based end-to-end methods but also is data efficient and robust to out-of-distribution examples. Moreover, it provides competitive performance in classifying segmented time series events with respect to 1NN-DTW with very cheap computational complexity.
We note that the approach assumes that the data being classified follows a certain generative model. Namely, each time series should contain an event with finite support, that is, with beginning and end within the recorded time series. We showed that for signal classes that do not contain events of finite duration the approach is less effective. 
Future work will include exploring ways to extend this method to classify time series events that are not readily segmentable.

\vspace{-1pt}
\section*{Acknowledgements}
This work was supported in part by NIH grant GM130825.
\vspace{-2.0pt}
\bibliographystyle{IEEEtran}
\bibliography{references}

\end{document}


\title{End-to-End Signal Classification in Signed Cumulative Distribution Transform Space}

\appendices
\section{Additional Facts and Remarks}

\noindent Observation 1: Let $c\neq p$ be two class labels, and  $\mathbb{S}^{(c)}$ and $\mathbb{S}^{(p)}$ be defined as in eq. (10), respectively. Assume that for any $\varphi^{(c)}_j\in \mathbb{S}^{(c)}$,  $f_{j,\varphi_w^{(p)}}\notin \textrm{span}\Big\{\big(\mathcal{G}_w^{(p)}\big)^{-1}\Big\}$ for any $w=1,...,M_p$ where $(f_{j,\varphi_w^{(p)}})^{\prime}\varphi_j^{(c)}(f_{j,\varphi_w}) = \varphi_w^{(p)}$, it follows that $\mathbb{\widehat V}^{(p)}_w\cap \mathbb{\widehat S}^{(c)}=\varnothing$ for any  $w=1,...,M_p$.

\noindent Proof. Assume by contradiction that there exists some $\varphi_j^{(c)}\in \mathbb{S}^{(c)}$ such that $\widehat \varphi_j^{(c)} \in \mathbb{\widehat V}^{(p)}_w\cap \mathbb{\widehat S}^{(c)}$. By the composition property and the definition of $f_{j,\varphi_w^{(p)}}$ above, we have that $ f_{j,\varphi_w^{(p)}}^{-1}\circ\widehat \varphi_j^{(c)} =  \widehat \varphi_w^{(p)}$ and hence $\widehat \varphi_j^{(c)} =  f_{j,\varphi_w^{(p)}}\circ \widehat \varphi_w^{(p)}$. Since by assumption $\widehat \varphi_j^{(c)}\in  \mathbb{\widehat V}^{(p)}_w $, it follows  that  $f_{j,\varphi_w^{(p)}}\in \textrm{span}\Big\{\big(\mathcal{G}_w^{(p)}\big)^{-1}\Big\}$, which is a contradiction.
\vspace{.2cm}

\noindent Remark 1: The last argument above follows from the simple fact that $ \mathbb{\widehat V}^{(p)}_w = \text{span}\Big(\mathbb{\widehat S}_{\varphi_w, \mathcal{G}_w^{(p)}}\Big) = \Big\{\sum\limits_{i=1}^k \beta_i f_{i,w}^{(p)}\circ \widehat \varphi_w^{(p)}: \beta_i\in \R\Big\}=\Big\{f\circ \widehat  \varphi_{w}^{(p)} : f\in \text{span}{\big(\mathcal{G}_w^{(p)}\big)^{-1}}\Big\}$. 
\vspace{.15cm}

\noindent Remark 2: By the disjointness assumption about different signal classes, i.e., $\mathbb{S}^{(p)}\cap \mathbb{S}^{(c)}=\varnothing$ for $c\neq p$, it follows that $f_{j,\varphi_w^{(p)}}\notin \big(\mathcal{G}_w^{(p)}\big)^{-1}$ for all $p=1,...,M_p$.  This property intuitively says that signals in $\mathbb{S}^{(c)}$ differs ``significantly" from the template $\varphi_w^{(p)}$ in the sense that they can not be recovered using   $\varphi_w^{(p)}$ under deformations  in $\mathcal{G}_w^{(p)}$. Recall that $\big(\mathcal{G}_w^{(p)}\big)^{-1} =\Big\{\sum\limits_{i=1}^k \alpha_i f_{i,w}^{(p)}: \alpha_i\geq 0\Big\} $, which differs from  $\text{span}\Big\{\big(\mathcal{G}_w^{(p)}\big)^{-1}\Big\}$ in the sense that the latter has no restriction on the $\alpha_i$'s. We comment loosely that the assumptions in Observation 1 is not unreasonable when signals in class $(c)$ are "significantly" different than templates in class $(p)$ and the deformation sets are reasonably ``small" in the sense that $f_{j,\varphi_w^{(p)}}\notin \textrm{span}\Big\{\big(\mathcal{G}_w^{(p)}\big)^{-1}\Big\}$. 


\section{Data Efficiency of the Proposed Method}\label{app:data_efficiency}
\begin{figure}[!htb]
     \centering
    \includegraphics[width=0.4\textwidth]{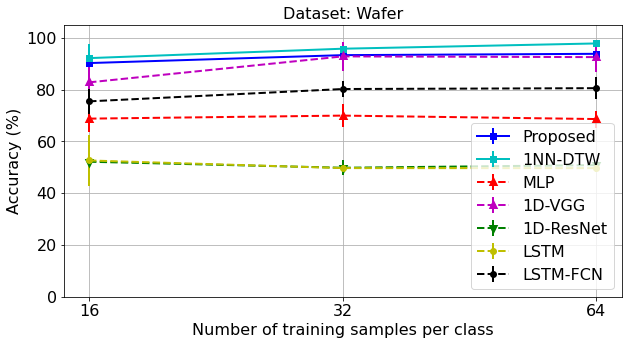}
    \includegraphics[width=0.4\textwidth]{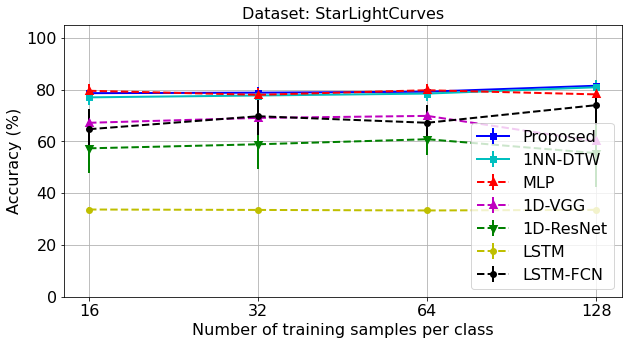}
    \includegraphics[width=0.4\textwidth]{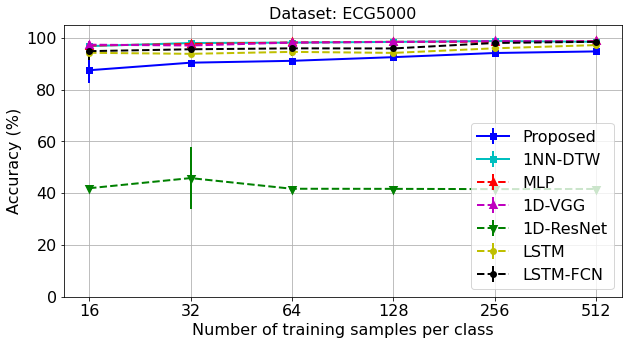}
    \includegraphics[width=0.4\textwidth]{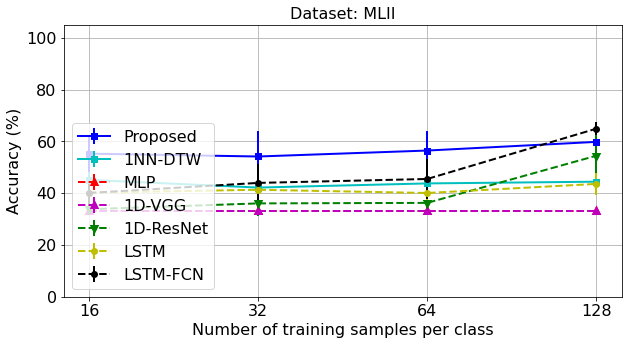}
    \caption{Accuracy as a function of number of training samples per class for different classification methods.}
    \label{fig:accVStrsamples_supp}
\end{figure}

Section IV-D of the paper shows that the proposed method is data efficient in comparison with the CNNs and provides two examples. Fig. \ref{fig:accVStrsamples_supp} shows the results from same experiment setup for four other datasets.
